\documentclass[12pt,a4]{article}
\usepackage[sectionbib, numbers, sort&compress, super, comma]{natbib}
\usepackage{color}
\usepackage{graphics}
\usepackage{subfig}
\usepackage{epstopdf} 
\usepackage[vmargin={25mm, 25mm},hmargin={25mm, 25mm}]{geometry}
\usepackage{graphicx, float}
\usepackage{mathtools}
\usepackage[normalem]{ulem}
\begin{document}
\setlength{\textwidth}{15truecm}
\setlength{\textheight}{23cm}
\bibliographystyle{unsrt}
\baselineskip=22pt
\bibliographystyle{unsrt}
%\setstcolor{red}

\begin{center}
{\Large \bf Vibrational Dynamics and Spectroscopy of Water at Porous 
 g-C$_{3}$N$_{4}$ and C$_{2}$N Materials} \\
 \vspace{1.0cm}
{ Deepak Ojha, Christopher Penschke and Peter Saalfrank$^{*}$}\\
Theoretische Chemie, Institut für Chemie, Universität Potsdam, Karl-Liebknecht-Strasse 24-25, D-14476 Potsdam-Golm, Germany \\
\end{center}

\begin{center}
{\bf Abstract}

\end{center}

%\baselineskip=18pt
%\begin{center}
Porous
 graphitic materials containing 
 nitrogen are promising catalysts 
 for photo(electro)\-chemical reactions, notably water splitting, 
 but can also serve as ``molecular sieves''.
 Hearby nitrogen, among other effects, increases the 
 hydrophilicity of the  
 graphite parent material.
 A deeper understanding of how water interacts 
 with C and N containing layered materials, if and which differences 
 exist between materials with different N content and pore size, 
 and what the role of water 
 dynamics are -- a prerequsite for catalysis and sieving, 
  is largely absent, however.
 Vibrational spectroscopy can answer some of these questions.
% From the theory side, vibrational fingerprints of water near 
% N-rich graphitic surfaces were so far  
% only analyzed by stationary calculations at low coverage and zero temperature.
 In this work, the vibrational dynamics and spectroscopy 
 of deuterated water molecules (D$_{2}$O) mimicking dense water layers at room temperature 
 on the surfaces of two different C/N based materials with different N content and pore size, namely 
 graphitic C$_{3}$N$_{4}$ (g-C$_{3}$N$_{4}$) and C$_{2}$N are studied using 
 Ab Initio Molecular Dynamics (AIMD). 
 In particular, Time-Dependent vibrational Sum-Frequency Generation spectra (TD-vSFG) of the OD modes
 and 
 also time-averaged 
 vSFG spectra and OD frequency distributions are computed. 
 This allows us to distinguish
 ``free'' (dangling) OD bonds from 
 OD bonds which are bound in a H-bonded water 
 network or at the surface -- with subtle differences between the 
 two surfaces and also to a pure  water / air interface. 
 It is found that 
 the temporal decay of OD modes is very similar on both surfaces with a
  correlation time near 4 ps. 
 In contrast, 
 TD-vSFG spectra reveal that the inter-conversion time 
 from ``bonded'' to ``free'' OD bonds being about
 8 ps for water on C$_{2}$N and thus twice as long 
 as  
 for g-C$_{3}$N$_{4}$, 
 demonstrating a propensity of the former material to stabilize 
 bonded OD bonds. 
%Also other subtle differences between the two catalysts are found, such as
% the different interaction  / orientation 
% of OD with either N or C sites on the surface. 
\noindent

$^{*}$E-mail: peter.saalfrank@uni-potsdam.de
\section{Introduction}
Water at interfaces and on surfaces is of immense significance for it plays a role  as a  solvent and as a catalyst in many chemical processes\cite{michaelides,marcus,skinner}. In biology, interfacial water at lipid bilayers and proteins contribute to the overall biological activity and selectivity\cite{pavel}. 
Water on semiconductor surfaces offers an alternative to conventional energy sources by means of electro- or photo-catalytic water splitting.
 Many  photo(electro)catalysts contain metals which are expensive and sometimes hazardous\cite{gonella,zhu,nicolas}. In contrast, recently synthesized organic materials like graphitic carbon nitride 
 (g-C$_{3}$N$_{4}$)\cite{anton1,anton2,anton3} and C$_{2}$N\cite{walczak2018,oschatz} not only have demonstrated their propensity to facilitate water-splitting, they are also easy to synthesize and non-hazardous to the environment.
These quasi-two dimensional, porous, layered materials offer a large surface area, and their properties can be tuned by the C/N ratio and the size of the pores. Nitrogen is particularly important in these materials, notably 
 for uptake of water, as it considerably increases  
 the hydrophilicity of the material w.r.t. graphite or graphene\cite{oschatz,penschke}.
 This makes these materials also interesting in a non-catalytic context, 
 {\em e.g.}, as sieves to separate liquids when acting as water-selective membranes.
\\ 

Both for catalytic and  non-catalytic purposes like selective adsorption and separation, a precise understanding 
 of the interaction of water with C/N-containing porous materials is  
 desirable.
Several theoretical studies have been performed which explored the geometric 
 and electronic structure of these carbon-based materials and how the latter 
 is modified by water adsorption\cite{wu,kasai,zou,peter,oschatz,penschke}. 
% While the adsorption of water on C/N-containing layered materials 
% both on the surface of and inside the material is possible\cite{oschatz,penschke}, most studies concentrate on% the surface, often considering single-layer models. 
 Often (but not always) single- or few-layer models were
 studied, and low coverages using stationary models. 
For example, in a first-principles Density Functional Theory (DFT) study 
 including van-der Waals corrections 
 and employing a (3$\times$3) periodic supercell model it was found, 
 that a single water molecule preferentially adsorbs in 
 a triangular pore of the g-C$_{3}$N$_{4}$ surface, with an adsorption energy 
 between -0.55 and  -0.62 eV depending on functional\cite{peter}. 
 The water molecule lies almost flat in the pore of the 
 buckled g-C$_{3}$N$_{4}$ surface, forming
 two hydrogen bonds with 
 accepting N atoms of the cavity. 
 Also cluster models using highly correlated wavefunction methods 
 support these results\cite{penschke,grueneis}.
 (Note that with smaller (1$\times$1) unit cells the 
 surface is falsely predicted to be flat and other 
 adsorption geometries are more stable\cite{kasai}.)
In Ref.\cite{peter}, by a Potential Energy Surface (PES) scan 
 also other (metastable) adsorption sites for 
 water were found, and barriers for lateral diffusion of water to 
 these sites were determined -- in the order of 
 0.2 eV. 
In Ref.\cite{zou}, in an attempt to go
 to higher-coverage situations, multi-layer systems 
  and dynamical behaviour relevant 
 for water-selective membranes and ``molecular sieves'',
 the diffusion of water (and ethanol) 
 through  pristine and modified g-C$_{3}$N$_{4}$ sheets was studied using classical molecular dynamics 
 simulations
 based on empirical forcefields. {An inverse correlation between the 
 diffusion coefficient of water and the lifetime of intermolecular 
 hydrogen bonds (HBs)
 was found, {\em i.e.} short lifetimes caused larger diffusion coefficients (and {\em vice versa}).}
 Diffusion coefficients (for water) on / between g-C$_{3}$N$_{4}$ were also found to be 
  smaller
  than for bulk water, and 
 the HB-lifetime was correspondingly larger for water at / in g-C$_{3}$N$_{4}$ 
 compared to bulk water\cite{zou}.  
%Furthermore, the hydrogen-bond lifetime which refers to the stability of hydrogen-bonds also shows an overall increase from 1.84 ps for the case of pure water {\color{blue}[Ref.?]} to 2.57 ps for water on the g-C$_{3}$N$_{4}$ surface\cite{zou}.} {\color{blue}[Where do you have this from?]}
\\

In case of C$_2$N, which consists of flat 
 layers under ideal conditions 
  and with hexagonal-shaped pores slightly larger than those of 
 g-C$_3$N$_4$, both periodic DFT and also cluster calculations 
 predict a similar adsorption energy for a single H$_2$O 
 on a C$_2$N monolayer as for g-C$_3$N$_4$.
 In this case, 
 however, the water resides above the (center of the) pore,  
 but still forming two H-bonds to two adjacent, basic N atoms\cite{oschatz,penschke}. Beyond single-layer models for C$_2$N and after loading with larger 
 amounts of water, according 
 to the combined experimental-theoretical work in Ref.\cite{oschatz}, 
 water molecules 
 reside between the layers, forming H-bonds with N atoms of C$_2$N which 
 are stronger than the H-bonds between different water molecules 
 in bulk water. The adsorption energy decreases with increasing water load.
\\

 In order to understand the interaction with and adsorption of liquid water on
 2D C/N-containing materials better, it is necessary to consider high coverages, finite temperatures, and dynamical behaviour which will naturally arise 
 at finite temperature. (In the studies mentioned above, this was only done within the classical MD
 work reported in Ref.\cite{zou}.) 
Our goal here is 
 to unravel differences of 
 the C/N material / water systems compared to bulk water or water / air interfaces, 
  and also differences between various C/N materials. 
 This is, in the end, a prerequisite to fully 
 understand the performance of 
 N-containing graphitic materials for water splitting or as sieves, 
 and their rational design.
 Another motivation 
 for our work is the mentioned 
 observation that the H-network 
 but also transport or chemical behaviour of water in confined spaces / at surfaces 
  can be very different from what is observed in the bulk (see also \cite{hummer,netz}).
 \\

Appropriate ``fingerprints'' 
 are needed by which differences between various systems 
 and general effects of a confining surface can be accessed.
 In this context, we note that 
 the spectral signature peaks for the concerted intermolecular motion of 
 hydrogen bonded water molecules as well as the intramolecular fluctuations 
 of the OH (or OD) modes of water molecules can be 
 seen in vibrational spectra, such as the infrared (IR) 
 spectrum.  In particular, the 
 IR spectrum provides a fingerprint region to decipher the hydrogen-bond dynamics in aqueous  solutions and at interfaces\cite{skinner}. Nevertheless, conventional vibrational spectroscopy cannot disseminate the contribution of interfacial molecules from the bulk. Vibrational Sum Frequency Generation Spectroscopy (vSFG) enables to selectively study the structure, hydrogen-bond network and orientational profile of interfaces and surfaces\cite{shen1, shen2, shen3, morita1, morita2, morita3, morita-book,tdk1}. Moreover, the temporal evolution of the interfacial molecules can be explored using Time-Dependent vSFG (TD-vSFG)\cite{tr-sfg,tdk2,tdk3}.  \\

Existing theoretical studies have not explored the vibrational dynamics of water molecules on two-dimensional, N-containing graphitic materials.
 In the present work, we have investigated the 
vibrational dynamics and vSFG spectra of interfacial water on g-C$_{3}$N$_{4}$ and C$_{2}$N surfaces using Ab Initio Molecular Dynamics. 
 We concentrate on surfaces in this work, using single-layer models 
 for the 2D material, and 
 leaving investigations of the effects of other layers and the bulk 
 to future work. We will consider thin layers 
 of D$_2$O molecules (rather than H$_2$O, 
 for practical reasons) adsorbed on one side of the 2D layer, and analyze 
 OD vibrations in detail.
 Fluctuations in the vibrational frequency of OD modes were obtained using the wavelet transform of the time-series analysis and the surface-specific Velocity Velocity Autocorrelation Function (ssVVAF) approach\cite{yuki} was used to calculate the vSFG and TD-vSFG spectra of interfacial water molecules. 
\\

The paper is organized as follows. In the following section \ref{sec2}, 
 we briefly outline theoretical methods and computational models. In Sec.\ref{sec3}, we
 study the vibrational dynamics, time-averaged vibrational spectroscopy 
 and time-resolved vibrational spectroscopy, respectively.
 Sec.\ref{sec4} summarizes and concludes our work.
\section{Computational Details and Models}
\label{sec2}
\subsection{Models}
We performed AIMD simulations for D$_2$O molecules adsorbed either 
 on C$_{2}$N or g-C$_{3}$N$_{4}$ monolayers, using periodic supercell models.
For g-C$_3$N$_4$, a supercell
of dimension ({20.4} $\times$20.4$\times$24.0) \AA \  with 126 atoms (C$_{54}$N$_{72}$) was employed representing a (3$\times3$) 2D trigonal elementary cell 
 to model the wave-like 
 reconstruction found elsewhere~\cite{peter,penschke} (see Fig.1(a) in Ref.\cite{penschke} for a pictorial representation). We then adsorbed 60 D$_2$O molecules on one side (the surface) of the monolayer. 
 A snapshot resulting from an AIMD simulation (see below), is shown in Fig.\ref{fig1}(a). 
 As a result, about 14 \AA \ of vacuum separate different periodic images along 
 $z$ (perpendicular to the surface).
% {\color{red}[Note: cos(pi/6)*20.4=17.67, while you use 18.0 -- why?]}{\color{green}[This should not be needed with revised dimensions (Christopher)]}.
\\

C$_{2}$N was modelled using a primitive unit cell of dimensions $(8.32 \times {8.32} \times 20.0)$ \AA \ with   
 18 atoms (composition C$_{12}$N$_6$), corresponding to a 2D trigonal elementary cell,  
 see Fig.1(c) in Ref.\cite{penschke}).
  On one side of this
layer, 12 D$_{2}$O molecules were added, as shown in the form of an AIMD snapshot in Fig.\ref{fig1}(b). This way, a vacuum gap of around 12 \AA \ remained between water-covered layers.
\begin{figure}
\begin{tabular}{cc}
(a) & (b)  \\
{\includegraphics[width=9cm, height= 4.8 cm]{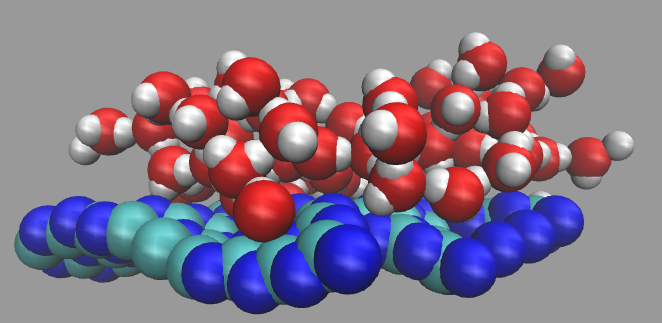}}
 &
{\includegraphics[width=6cm, height=4.8 cm]{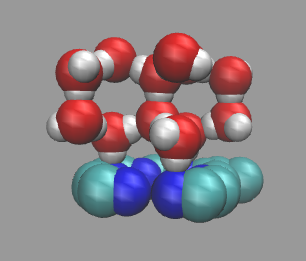} }
\end{tabular}
\caption{Snapshots of water (D$_2$O) molecules simulated on (a) a C$_{3}$N$_{4}$ 
 (3$\times$3) monolayer and (b) a C$_{2}$N monolayer. The employed unit cells 
 are shown in side-views, with 
 C: green, N: blue, O: red, D: white. 
%{\color{red}[Are these rectangular or trigonal unit cells?]}}{\color{green}[ Trigonal ]
}
\label{fig1}
\end{figure}
\subsection{AIMD simulations}
First-principles simulations of water molecules on the g-C$_{3}$N$_{4}$ and C$_{2}$N surfaces  were performed using the Vienna Ab initio Simulation Package\cite{vasp1,vasp2} (VASP, version 5.4). The inner core electrons were represented by
 projector augmented-wave (PAW) pseudopotentials\cite{psp,blochl} and valence electrons  using the Perdew Becke Ernzerhof (PBE) exchange correlation
functional\cite{pbe} 
 together with Grimme's D3 dispersion correction\cite{dftd, dftd2}. 
 (For a general assessment of DFT for water, see Ref.~\cite{michal2}.)
The plane-wave kinetic energy cutoff was set to 400 eV for 
 both systems. For C$_2$N/water, a 3$\times$3$\times$1 k-point grid was used 
 for Brillouin zone sampling; for g-C$_{3}$N$_{4}$/water,  
 only the $\Gamma$-point was included.
\\

Simulations were
performed in the NVT ensemble using the  Nos{\'e}-Hoover thermostat\cite{nose} at 300 K with deuterium masses for hydrogen and a time step of 1 fs for integrating the equations of motion. 
The vibrational spectra were sampled by using single, long
 NVT trajectories of 45 ps length (C$_3$N$_4$/D$_2$O) and 
 55 ps (C$_2$N/D$_2$O), respectively.
 D$_2$O instead of H$_2$O was used in order to allow for longer timesteps and total propagation times in AIMD, hence better statistics, and also to inspire experiments which are often done with deuterated water~\cite{melani2}.
\subsection{Frequency distributions}
A (time-dependent) distribution of 
 vibrational frequencies 
 of OD modes of  water molecules on g-C$_{3}$N$_{4}$ and C$_{2}$N surfaces 
%corresponding to the $0\rightarrow 1$ vibrational level excitation,
%\begin{equation}
%\delta\omega(\tau) = \omega(\tau) - \left \langle \omega \right \rangle
%\end{equation}
% with $\left \langle \omega \right \rangle$ being the average vibrational frequency for the simulated trajectory {\color{red}[1. From where do you get $\omega$ for a given OD bond at time $t$? 
% 2. Why do you call this $0 \rigtharrow 1$ and $10$? This is a {\em classical} calculation! 
%2. You run only a single trajectory for 50 ps, or do you 
% run several?]},  
% $P(\omega)$?]}, 
 was  determined using the wavelet transform of a time-series analysis. 
The method is well 
 documented\cite{wavelet} and 
 is based on the principle that a time-dependent function ($f(t)$) can be expressed in terms of basis functions obtained by the translations and dilations of a mother wavelet
\begin{equation} \label{eq1}
\psi_{a,  b}(t)=a^{-\frac{1}{2}}\psi\Bigg(\frac{t-b}{a}\Bigg) \quad ,
\end{equation}
which is represented in the so-called Morlet-Grossman form in our present study and is mathematically given as,
\begin{equation}
\psi(t) = \frac{1}{\sigma \sqrt{2 \pi}} e^{2\pi i\lambda t}e^{-\frac{t^{2}}{2\sigma ^{2}}} \quad .    
\end{equation}
The parameters $\lambda$ and $\sigma$ are assigned the values of 1 and 2 sec$^{-1}$ in our present study. 
 The coefficients of the wavelet expansion are given by the wavelet transform of $f(t)$, 
 {\em i.e.}
\begin{equation} \label{Eq4}
    L_{\psi}f(a,  b)=a^{-\frac{1}{2}}\int_{-\infty}^{+\infty}f(t){\psi} 
\Bigg(\frac{t-b}{a}\Bigg)dt \quad , 
\end{equation}
where $a$ and $b$ are both real quantities.
Here, $a>0$ is a scale parameter which is directly related to the instantaneous frequency content of the system over a small time-window centered around $b$. Accordingly, based on the fluctuations in the time-series $f(t)$, the wavelet transform $L_{\psi}f(a,  b)$ provides the vibrational frequency for the given small time-window around $t=b$. 
%Further, for a wavelet $\psi$ with  center   $t^{*}$ and radius $\Delta_{\psi }$,  $L_{\psi}f(a,  b)$ localizes the function within  the time window of width B, i.e. $\left [ b + at^{*}-a\Delta_{\psi }, b + at^{*}+a\Delta_{\psi } \right ]$.  
 Since the frequency value is proportional to $\frac{1}{a}$, the time-window narrows for high frequency (small $a$) and widens for low frequency (large $a$). 
%On similar line, for a wavelet with its Fourier transform $\widehat{\psi}$ and central frequency $\omega^{*}$,
%the frequency fluctuations are obtained within the domain A i.e. $\left [ \frac{\omega^{*} }{a}-\frac{\Delta_{\widehat{\psi} } }{a}, \frac{\omega^{*} }{a}+\frac{\Delta_{\widehat{\psi } } }{a} \right ]$. 
%The wavelet transform $L_{\psi}f(a,  b)$, localizes the function $f$ in a time-frequency window.
% of $A \times B$.
The value of parameter $a$ which maximizes the modulus of the wavelet transform 
of time series $f$ at time $t=b$ is used to calculate the most important frequency component for the given interval.
\\

The time-series of interest in this work, $f(t)$, is constructed as a complex function with its real and imaginary parts corresponding to fluctuations 
 of the bond length and momentum of an OD mode projected along the OD bond, namely 
 (possible unit factors omitted)
\begin{equation} \label{Eq4}
f(t) = \delta r^{\mathrm{OD}}(t) + i \delta p^{\mathrm{OD}}(t) \quad . 
\end{equation}
 Here, the fluctuations 
 are defined as 
\begin{equation}
\delta q(t) = q(t) - \left \langle q \right \rangle
\label{eqdelta}
\end{equation}
where $q(t)$ denotes the instantaneous property 
 ($=r^{\mathrm{OD}}(t)$ or $p^{\mathrm{OD}}(t)$), and 
 $\left \langle q \right \rangle$ the corresponding average 
 over the trajectory.
% (or several trajectories).
% {\color{red}[Is this correct? I mean, are these fluctuations or just 
% instantaneous values?]}{\color{green} [Yes this is mathematically correct]} 
 Here, 
 $r^{\mathrm{OD}}(t)= 
|\underline{r}^{\mathrm{OD}}(t)|$ where $\underline{r}^{\mathrm{OD}}(t)=\underline{r}_\mathrm{O}-\underline{r}_\mathrm{D}$ is the OD bond vector for a given OD bond, 
  obtained from atomic position vectors. Further, 
 the  momentum projected along the OD mode is given as
\begin{equation} 
p^{\mathrm{OD}}(t) = \left(\frac{m_{\mathrm{D}} \cdot m_{\mathrm{O}}}{m_{\mathrm{D}}+m_{\mathrm{O}}}\right) \cdot (\underline{v}_{\mathrm{O}}-\underline{v}_{\mathrm{D}}) \cdot \underline{\hat r}^{\mathrm{OD}} \quad ,
\end{equation}
where $m_{\mathrm{D}}, m_{\mathrm{O}}$ are atomic masses of deuterium and 
 oxygen, 
${\underline{v}}_{\mathrm{O}}$, ${\underline{v}}_{\mathrm{D}}$ are atom velocities, and 
 ${\underline{\hat{r}}}^{\mathrm{OD}}={\underline{r}}^{\mathrm{OD}}/r^{\mathrm{OD}}$ the unit vector along the OD bond. This method is then applied to all the OD modes present in a given system. 
For further details and recent 
 applications of the method within AIMD, see Refs.\cite{wavelet, tdk4, tdk5, ojha}.
%{\color{red}[Check all carefully. ]} {\color{green} [ all references are correct]}
%
\section{Results}
\label{sec3}
\subsection{Vibrational frequency distribution and dynamics}
\label{sec31}
\subsubsection{Time-averaged frequency distribution}
 We first look at time-averaged vibrational frequency distributions 
 $P(\omega)$ of OD modes, obtained from time-averaging 
 the time-dependent vibrational frequencies of OD modes 
 resulting from the wavelet transform of $f(t)$.
% {\color{red}[How precisely was $P(\omega)$ computed (formula, $P(\omega)= ...$)? What is the scale in the figures below?] }
 {Practically,  
 an averaged vibrational frequency distribution is obtained by dividing the frequency range of 2000-3000 cm$^{-1}$
 in  equally spaced bins. If at a given instant the frequency of an OD mode falls within a given bin, the bin height is 
 incremented by unity. This process is repeated for each OD mode and along the entire  trajectory.}
\\

\begin{figure}[!htb]
\vspace*{-1cm}
\begin{center}
\includegraphics[height=12cm]{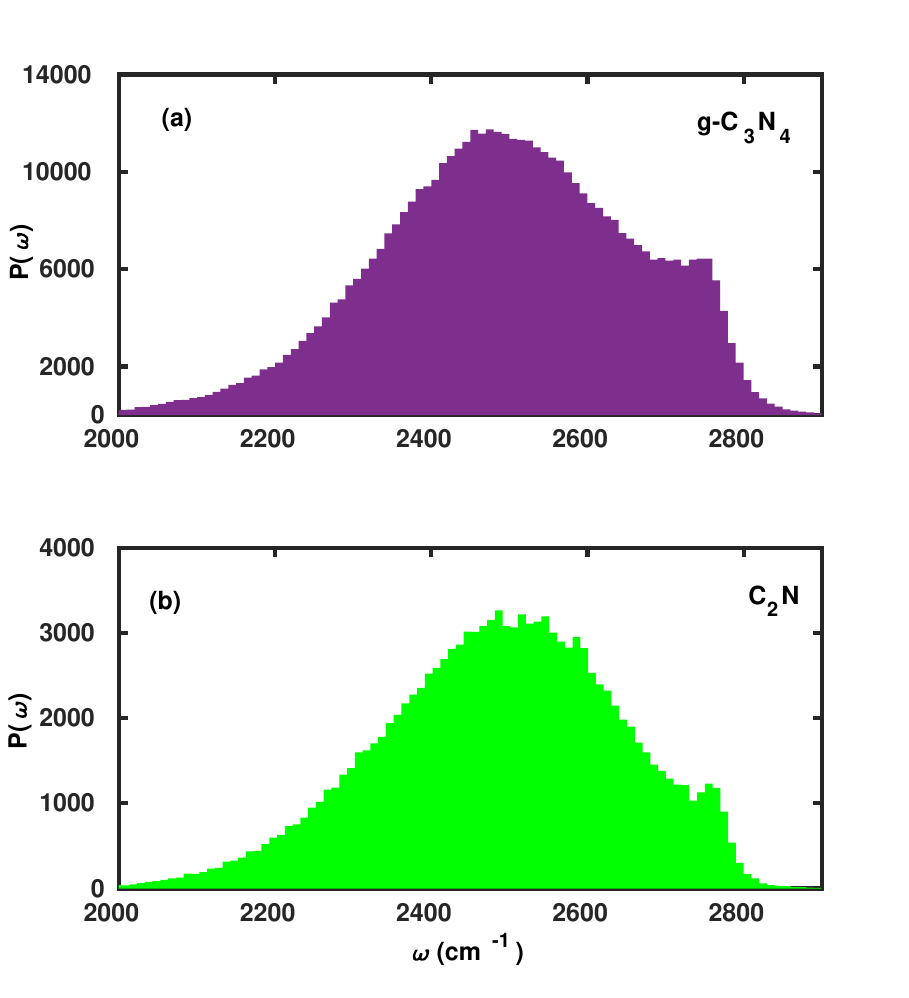}
\end{center}
\caption{\label{fig2} Time-averaged frequency distribution $P(\omega)$
 (in arbitrary units) 
 of the OD modes of water molecules on the (a) g-C$_{3}$N$_{4}$ and (b) C$_{2}$N surface.
 } \end{figure}

The time-averaged vibrational frequency distribution
 of OD modes of the interfacial water on
 the g-C$_{3}$N$_{4}$ surface was found to be a distribution
 ranging from about 2000 cm$^{-1}$
 to about 2900 cm$^{-1}$ with a
 mean at 2496 cm$^{-1}$ as shown in Fig.\ref{fig2}(a).
 The distribution is similar to a distribution found 
 for bulk water at 300 K using a similar methodology (see Fig.2(a) of Ref.\cite{ojha}, for example), 
 however, with two differences. First, the distribution in our Fig.\ref{fig2}(a)
 is slightly broader, and, more importantly, a clear shoulder 
 peak now appears in the high frequency region around 2750 cm$^{-1}$ which 
 is absent in bulk water. Closer analysis shows that this 
 shoulder corresponds to the free or dangling OD modes, {\em i.e.}, 
 OD bonds which are not part of a H-bond network and not bound to the surface 
 and stick out into vacuum instead.
Similarly, the average vibrational frequency of OD modes on the C$_{2}$N surface is 2487 cm$^{-1}$ (Fig.\ref{fig2}(b)), and the 
 whole time-averaged distribution is similar to the other surface.
 Also for C$_{2}$N, a shoulder peak in the  high-frequency region around 2750 cm$^{-1}$ 
 is found, again mainly due to the free/dangling OD modes of the interfacial water. Due to the non-centrosymmetric environment at the surface, the 
 tetrahedral hydrogen-bond network is distorted for both surfaces 
 and thus the population of free OD modes shows an overall increment. The high-frequency shoulder is more prominently observable for g-C$_{3}$N$_{4}$ which implies comparatively higher population of free/dangling OD modes. Interestingly,
  high-frequency shoulders are also 
 a characteristic feature of vSFG (vibrational Sum Frequency Generation) spectra of other systems with interfacial water\cite{morita1,morita2,morita3,melani,melani2}. 
\subsubsection{Time-resolved frequency distribution and dynamics}
Further, the temporal dynamics of the OD modes of water molecules on the surface of g-C$_{3}$N$_{4}$ and C$_{2}$N were studied using time-dependent joint probability distributions. 
The time-dependent joint probability distribution used 
 can be expressed mathematically as
\begin{equation}
P(\omega_{3},t_{2},\omega_{1}) = \left \langle \delta\left ( \omega(t_{2})-\omega_{3}   \right ) \cdot \delta \left ( \omega(0)-\omega_{1}   \right )  \right \rangle \quad .
\end{equation}
Here, $\delta(a(t)-b) = a(t)-b$ in analogy to Eq.(\ref{eqdelta}), and the average 
 is taken over all OD bonds.
%{\color{red}[Is that correct?  Also the statement about the average? 
%And what was time-zero?]}
% To calculate distribution, 
% in practice the trajectory was divided into small segments with length at 
% least equal to (or larger than) time $t_{2}$, 
% and the rigin zero is vibrational frequency at the first time instant of the  trajectory segment]}
The joint probability distribution gives
 the probability that a given OD mode which was oscillating with 
 the vibrational frequency $\omega_{1}$ evolves to a
 frequency $\omega_{3}$ within a time interval $t_{2}$. 
 (Note that for centrosymmetric 
 systems other than those studied here,  
 $P(\omega_{3},t_{2},\omega_{1})$ is similar to 
 a two-dimensional IR (2D-IR) spectrum at long waiting times 
 $t_2$~\cite{ojha}.) 
\\
\begin{figure}[!htpb]
\vspace*{-1cm}
\begin{center}
\hspace*{-1cm} \includegraphics[width=1.3\textwidth]{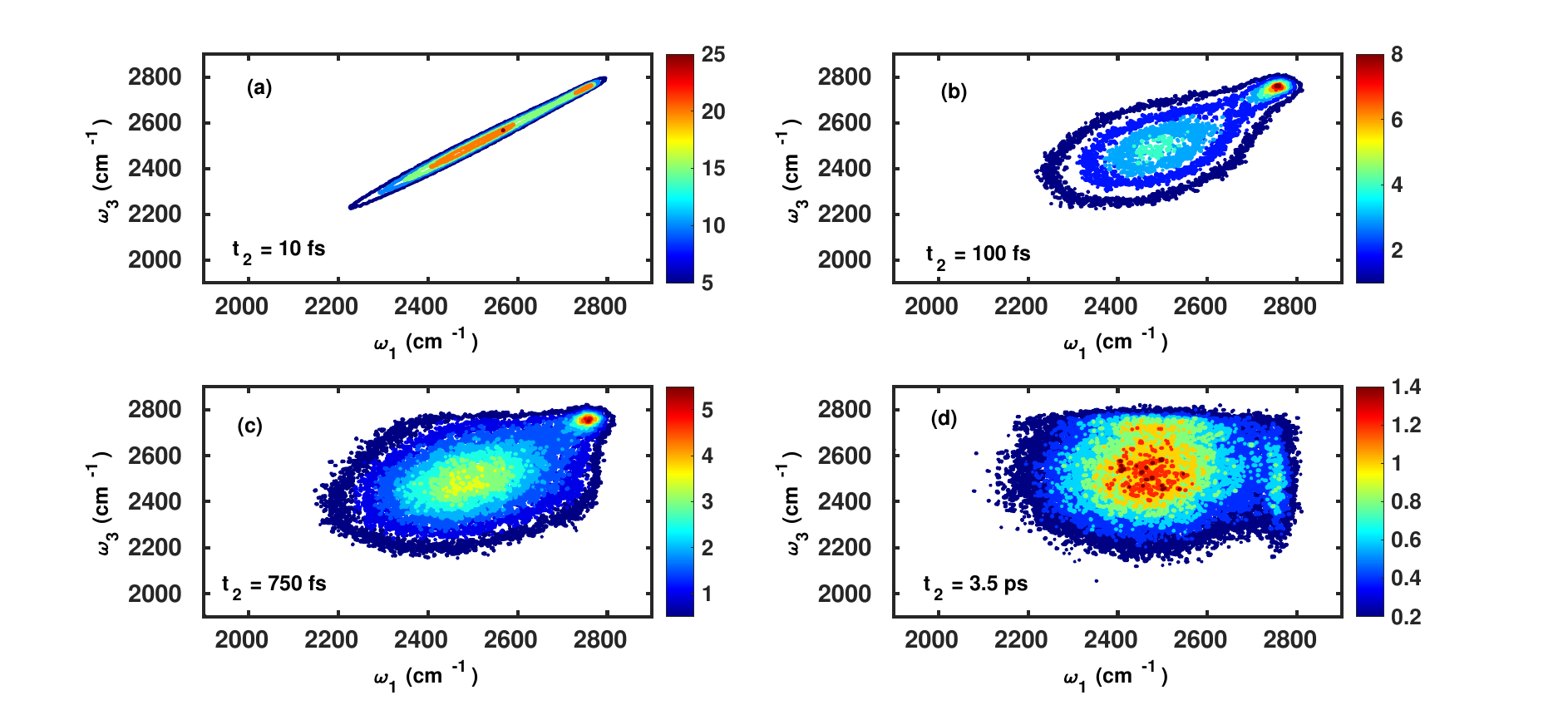}
\end{center}
\caption{\label{fig3} Joint probability frequency distribution 
 $P(\omega_{3},t_{2},\omega_{1})$ of OD modes of  water molecules on the  g-C$_{3}$N$_{4}$ surface  for waiting times $t_{2} = 10, 100, 750 $ and $3500$ fs.
% {\color{red}[Give a colour code.]} {\color{green}[ colour code is attached with the each distribution.]}
} \end{figure}

In Fig.\ref{fig3}, we have plotted the distributions for the waiting times $t_{2} = 10, 100, 750  $ and $3500$ fs for OD modes of water molecules on the g-C$_{3}$N$_{4}$ surface respectively.  For small waiting times, {\em i.e.},  $t_{2} = 10$ fs, the distribution is predominantly a straight line aligned along the diagonal $\omega_1 = \omega_3$, indicating that the OD modes keep close to their 
 initial vibrational frequency. However, for the longer waiting times, the distribution eventually  gets elongated and evolves into a  spherical distribution which implies that the OD modes undergo vibrational spectral diffusion due to hydrogen bond rearrangement. Moreover, within 3.5 ps the OD modes have sampled all accessible vibrational states / frequencies ({\em cf.} Fig.2(a)). A key observation in the time-resolved frequency distributions is that the high frequency shoulder is evidently observable at low as well as long waiting times. Further,  the vibrational dephasing of the modes corresponding to the shoulder peak is seen progressing as vertical slab along 2750 cm$^{-1}$.
\\

\begin{figure}[!htpb]
\vspace*{-1cm}
\begin{center}
\hspace*{-1cm} \includegraphics[width=1.3\textwidth]{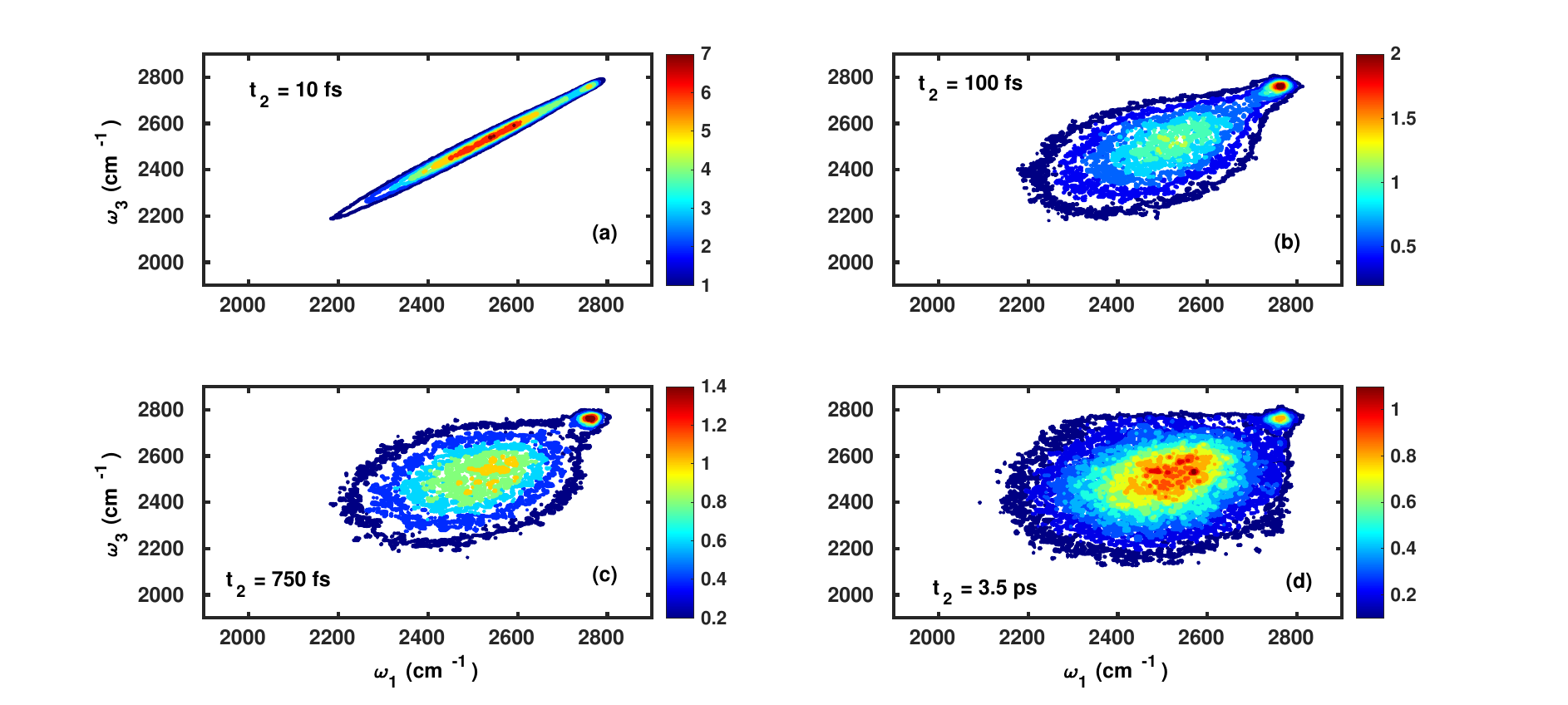}
\end{center}
\caption{\label{fig4}Joint probability frequency distribution 
 $P(\omega_{3},t_{2},\omega_{1})$ of OD modes of water molecules on the C$_{2}$N surface for waiting times $t_{2} = 0, 100, 750 $ and $3500$ fs.
%{\color{red}[Give a colour code.]} {\color{green}[color code is provided]}
} \end{figure}

A similar analysis for the water molecules on the C$_{2}$N surface corresponding to the waiting times of $t_{2} = 10, 100, 750  $ and $3500$ fs is shown in Fig.\ref{fig4}. 
 The vibrational dephasing follows a similar pattern, and 
 overall differences between 
 both surfaces seem small. 
\\

\begin{figure}[!htpb]
\begin{center}
%\begin{tabular}{cc}
%(a) & (b) \\
\hspace*{-2cm}
\includegraphics[height=8cm,width=20cm]{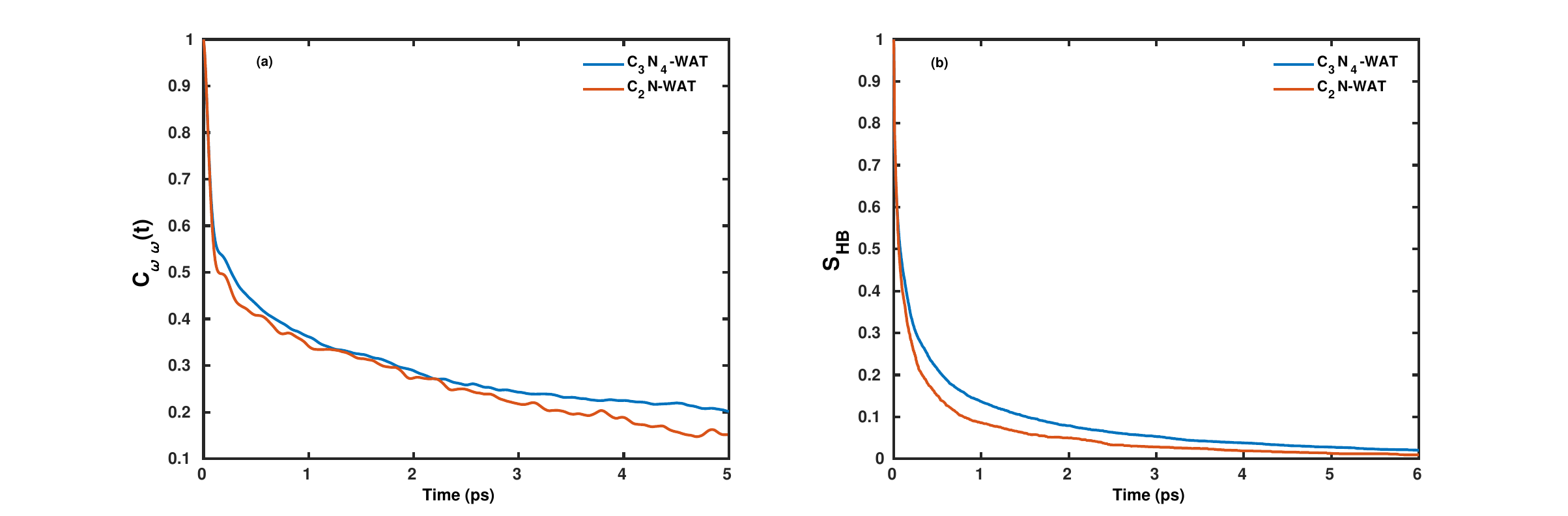}
%&
%\hspace*{-1cm} 
%\includegraphics[height=8cm,width=10cm]{shb.eps}
%\end{tabular}
\end{center}
\caption{\label{fig5} (a) Frequency correlation 
 function $C_{\omega \omega}(t)$ and 
 (b) hydrogen bond correlation function $S_{HB}(t)$ for OD modes of water molecules on the g-C$_{3}$N$_{4}$ and C$_{2}$N surfaces.
%{\color{red}[Redo figures (since this is no one figure only)?]} {\color{green}[figure revised]}
} \end{figure}

To more quantitatively address this point, 
the temporal dynamics of the OD modes of water molecules  on  the g-C$_{3}$N$_{4}$ and C$_{2}$N surfaces is also analyzed using a frequency time correlation function (FTCF) which is given
as,
\begin{equation}
C_{\omega \omega}(t) = 
\frac{\left \langle \delta \omega(0)\cdot \delta \omega (t)  \right \rangle}{\left<  \delta \omega \left ( 0 \right ) ^{2}\right>} \quad .
\end{equation}
%Similar to the evaluation of $P(\omega_{3},t_{2},\omega_{1})$ above, averages are  
% over the entire frequency trajectory of all the OD modes. The calculation involves division of the trajectory in small segments and then the product as shown in the numerator is calculated. 
% The vibrational frequency ot the beginning of each segment is labelled as $t=0$.
% {\color{red}[I (still) do not know what that means. A frequency is not a time.]}
The time-dependent decay of the correlation function 
 of OD modes of water molecules on the g-C$_{3}$N$_{4}$ and  C$_{2}$N surfaces 
 is shown in Fig.\ref{fig5}(a).
\\

Evidently, the decay is biphasic with short time regime which extends up to 100 fs and is governed by the ultrafast librational dynamics or concerted collective motion of water molecules\cite{marx} whereas the long time decay which extends up to a few picoseconds and is mainly attributed to the hydrogen bond rearrangement dynamics. The timescale of vibrational dephasing was determined by using a bi-expontential fitting function,
\begin{equation}
 C_{\omega \omega}(t) = a_{0} \exp\left(-\frac{t}{\tau _{0}}\right) + (1-a_{0}) \exp\left(-\frac{t}{\tau _{1}}\right) \quad .
\end{equation}
The short time constant $\tau_{0}$ was found to be 76 fs and the long time decay constant $\tau_{1}$ was 5.7 ps for water on g-C$_{3}$N$_{4}$.
 For C$_2$N, we find $\tau_0=104$ fs and $\tau_1=4.4$ ps.  
 We note that for both systems the long time constant,
 $\tau_1$,  
 is significantly longer than for
 bulk liquid water\cite{chandra,ojha}.
 According to Ref.~\cite{ojha}, a
 computed $\tau_1$ for bulk D$_2$O is 1.6 ps. 
 Thus, vibrational dynamics of OD modes are slowed down in proximity of a 
 the g-C$_{3}$N$_{4}$ surface, and slightly less so near C$_2$N.
\\

Since frequency fluctuations of the OD modes are
 associated with the rearrangement of the local hydrogen bond network, we have also calculated a ``hydrogen bond lifetime'' using a continuous HB correlation function given as, 
\begin{equation}
S_{HB}(t) = \frac{\left<h(0)H(t) \right>}{\left< h(0)^2 \right>} \quad . 
\end{equation}
Here, $h(t)$ and $H(t)$ are hydrogen bond population labels, defined as follows.
  For a given pair of hydrogen-bonded, different
  water molecules at an instant $t$, $h(t)$ is assigned a value of 1 and  otherwise 0. 
 (Time zero refers to the beginning of the segment of the 
 trajectory into which the trajectory was divided.)
% {\color{red}[There is no h(t), only h(0), so h is always the same, at least if we have a single trajectory only. Also, what is the average taken over if there is a single trajectory only? Over the various OD connections? Which of the OD connections are taken into account?]} {\color{green}[For the calculation again the trajectory is divided in segments and the origin(t=0) is also moved systematically so h(0) can change value.  Moreover, a simple distance cutoff is used to identify intermoleuclar hydrogen bonds. ]} 
Further, if the hydrogen bond between a pair of water molecules remains intact from time $t=0$ till time $t$, $H(t)$ is assigned a value of 1 or else 0. The water molecules are considered to be hydrogen-bonded, if the intermolecular O $\dots$ D distance is less than 2.45 \AA \ corresponding to the first minimum in the O $\cdots$ D pair correlation function. 
% {\color{red}[Don't we also need a lower boundary (to exclude covalent O-D bonds?)]} {\color{green}[ We are looking only for the intermolecular hydrogen bonds between different water molecules. So for the entire calculation the distance between O and D is calculated only if they belong to different water molecules]}  
 The temporal decay of the hydrogen bond correlation function 
 indicates the timescale 
 of rearrangement of the H-bond network 
 and is shown for water (D$_2$O) on g-C$_{3}$N$_{4}$ and C$_{2}$N surfaces in 
 Fig.\ref{fig5}(b).
Evidently, also within this measure, the temporal decay of the 
 $S_{HB}(t)$ correlation function is somewhat faster for 
 C$_{2}$N/water in comparison to g-C$_{3}$N$_{4}$/water, 
 as can be seen by visual 
 inspection of Fig.\ref{fig5}(b). 
 More quantitatively,  the long hydrogen bond lifetime
 $\tau_1$,  obtained from bi-exponential fits 
 of $S_{HB}(t)$, 
 are found to be $\tau_1=1.8$ ps for g-C$_{3}$N$_{4}$ 
 and 1.3 ps for water on the C$_{2}$N surface, respectively.
% {\color{red}[The difference is still small, but looks much bigger 
% in the figure. Can you check the fit again?]} {\color{green}[I have revised the calculations and based on them new lifetime for C3N4 is 1.8 ps and for C2N is 1.3 ps]}.
% Since, the $S_{HB}$ is defined by using distance based hydrogen-bond criteria, the ultrafast intermoleculear dynamics cannot be characterized within its formulation. As a result,  only the long-time component is provided here which can be mapped to the hydrogen-bond lifetime. 
 Calculations of $S_{HB}(t)$ have been done also for bulk deuterated 
 water and for 
 the water-vacuum interface in Ref.\cite{chandra2}. The H-bond lifetimes 
 computed there are of similar length, but hard to compare quantitatively to 
 our numbers since they were obtained with another DFT functional 
 and with Car-Parrinello MD rather than Born-Oppenheimer MD. Still, 
 in Ref.~\cite{chandra2} it was found that the 
 bulk $S_{HB}$ function decays faster than the 
 interface one. Here, we have the additional effect of the 
 porous surface and we see differences between the two surfaces, 
 g-C$_{3}$N$_{4}$ and C$_{2}$N.
\subsection{Vibrational Sum Frequency Generation (vSFG) spectra}
\label{sec32}
Further insight into the water dynamics on the studied 2D materials, 
 more directly accessible by experiment, comes from vibrational 
 spectroscopy. Here we 
 concentrate on vibrational SFG  as a particularly surface-sensitive 
 method.
\subsubsection{Time-averaged vSFG spectra}
We start with time-averaged vSFG spectra.
A time-averaged vSFG spectrum provides relevant 
 frequencies and gives information about the orientation profile of molecular dipoles at the surface/interface. The time-averaged vSFG spectrum 
 is determined 
 from the second-order 
 optical susceptibility tensor of 
 the system, with tensor elements\cite{morita-book,tdk1}
\begin{equation} \label{eq2}
{\chi^{2}_{abc}} \propto  \int e^{-i \omega t} \left \langle  {{ { \alpha}_{ab}}(t) { { { \mu}_{c}}}(0)}  \right \rangle 
 dt \quad .
\end{equation}
Here, $a$, $b$, and $c$ are Cartesian coordinates, $\alpha_{ab} (t)$ 
 is the ($a$,$b$)-component of the polarizability tensor at time $t$, 
 $\mu_c(0)$ the $c$-component of the dipole vector at time $t=0$, and $\omega$ the frequency.
 Squared, summed (and weighted)
 components of ${\chi^{2}_{abc}}$ give the vSFG spectrum.
 Here, we concentrate on so-called ssp-polarization 
 spectra, which we approximate by 
 only considering a single 
 ${\chi^{2}_{abc}}$ element and setting $a=b=x$ and 
 $c=z$, respectively, where $x$ is a direction parallel to the surface and $z$ perpendicular to it. We show real and imaginary parts 
 of the second-order susceptibility, 
 ${\chi^{2}_{xxz}}(\omega)$, 
 whereby the imaginary part can be 
  attributed to the orientation of the interfacial molecules with respect to the surface\cite{shen1,shen2,shen3}. 
 In practice, we calculate here 
the ssVVAF method mentioned earlier\cite{yuki,melani}. Then, 
 Eq.(\ref{eq2}) can be rewritten, for our D$_2$O / surface vSFG problem 
 (with only O-D stretch vibrations considered), as\cite{yuki,melani} 
\begin{equation}
 {\chi^{2}_{abc}} (\omega)  \propto
\begin{dcases}
    \dfrac{\mu'_{\mathrm{OD}}\alpha'_{\mathrm{OD}}}{i\omega^2} \int^{\infty}_{0}dt \ e^{-i\omega t}   \left \langle \sum_{i,j}^M  {{v}}^{\text{OD}}_{c,i} (0) \dfrac{{\underline{v}}^{{\text{OD}}}_j (t) \cdot {{\underline{r}}}^{{\text{OD}}}_j (t) }{ |{\underline{r}}^{\text{OD}}_j (t)|}   \right \rangle , &  a = b \\     0,              & a\neq b \quad .
\end{dcases}
\label{eq3}
\end{equation}
Here, $\underline{r}^{{\text{OD}}}_i$ is 
 the coordinate vector of an OD bond ($i$) as defined earlier 
 (there are $M=2 \cdot N$ O-D bonds for $N$ water molecules), 
 $\underline{v}^{{\text{OD}}}_{i} = d \underline{r}^{{\text{OD}}}_{i}/dt$ 
 the corresponding velocity, ${{v}}^{\text{OD}}_{c,i}$ its $c$-component, 
 and $\mu'_{\mathrm{OD}}$ and $\alpha'_{\mathrm{OD}}$ are derivatives 
 of the dipole moments and polarizabilities, respectively. 
 A few things are worth mentioning:
 The ssVVAF expression (\ref{eq3}) is approximate in the sense that 
 only O-D stretch vibrations are considered. Further, we also use a 
parameterized (linear) version of dipole moments and polarizabilities without 
 explicitly calculating them -- in fact we also 
 do also not include the frequency dependence of $\mu'_{\mathrm{OD}}$ and $\alpha'_{\mathrm{OD}}$ and 
 set them as constants. Finally, 
 a sometimes used ``quantum correction factor'' $Q(\omega)$\cite{yuki} is also neglected. 
 Still, the approach followed here will give insight into the O-D dynamics and in particular 
 into the dynamics of the interfacial H-bond network.
 The method has been shown to give for the 
 O-H stretch region of a water / air interface, quite similar vSFG spectra 
 compared to a non-parameterized procedure with calculated dipole moments and 
 polarizabilities\cite{tdk1}.
\\

\begin{figure}[!htb]
\begin{center}
\includegraphics[width=1.2\textwidth]{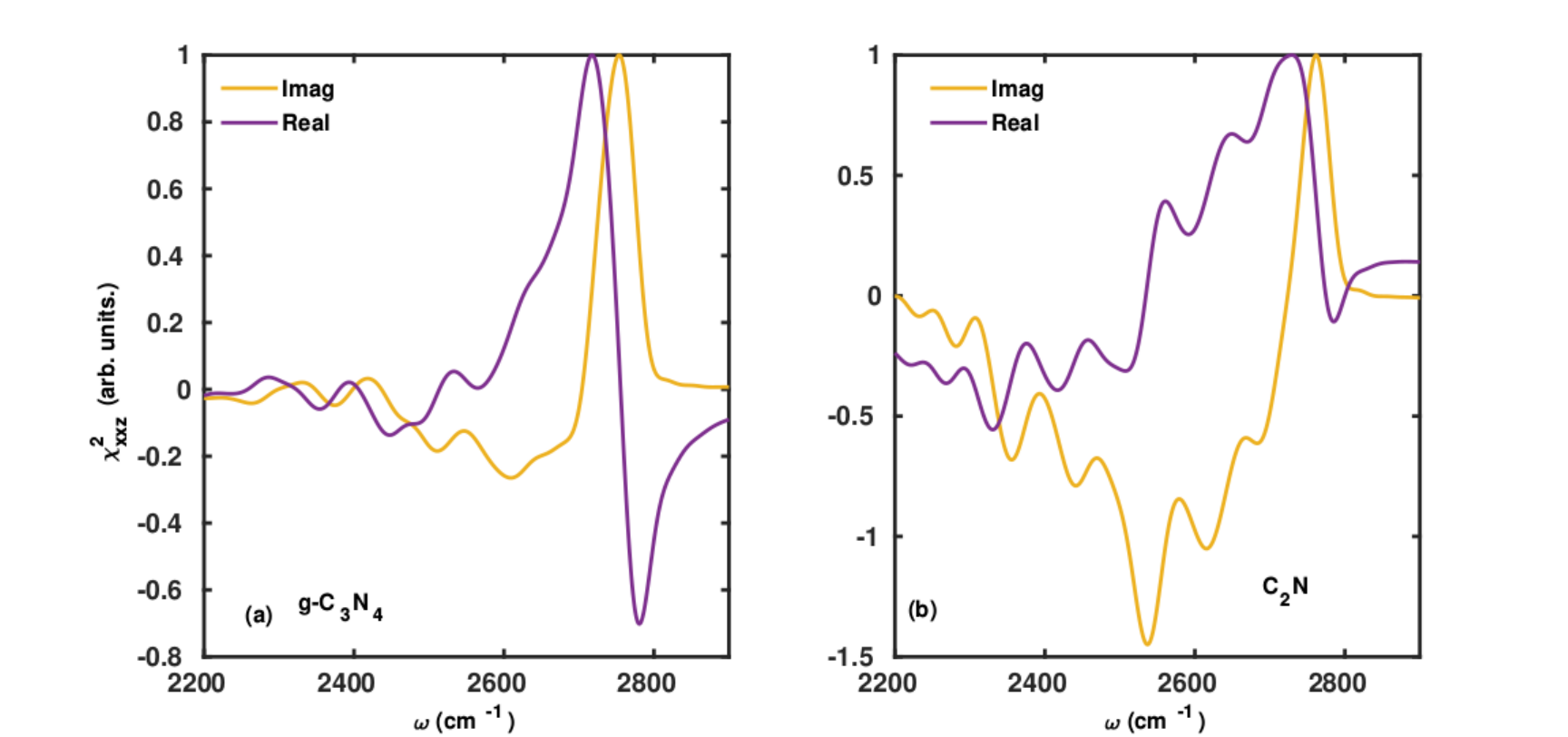}
\end{center}
\caption{\label{fig7} 
 Real and imaginary parts of $\chi^{2}_{xxz}$ 
 for D$_2$O at (a) g-C$_3$N$_4$ and (b) C$_2$N surfaces. The y-axis 
 is in arbitrary units, with the largest positive intensity of 
 imaginary or real parts normalized to 1.
 } \end{figure}

In Fig.\ref{fig7}, we show the real and imaginary parts of $\chi^{2}_{xxz}$
 for D$_2$O at g-C$_3$N$_4$ (a) and C$_2$N, respectively, again
 in the frequency range of the OD stretch vibrations. 
%We first of all note that the largest real and imaginary parts of the second-order susceptibility 
% are found around $\sim 2750$ cm$^{-1}$, characteristic for free or dangling OD bonds, {\em i.e.}, those not 
% bound in a H-bond network as mentioned earlier.
 To gain information on the orientation of OD bonds,  
 we focus on the imaginary part of 
 $\chi^{2}_{xxz}$.
(The real part of the vSFG spectrum is similar to a 
 conventional infrared spectrum and provides information about the frequency required to excite vibrations of the adsorbed molecules.) 
For D$_2$O/g-C$_3$N$_4$, the imaginary 
  part of the susceptibility is characterized by a broad, (mostly)
  negative peak for the frequency domain of 2400-2700 cm$^{-1}$ which refers to the OD modes which are hydrogen-bonded with the other water molecules or with the g-C$_{3}$N$_{4}$ surface (Fig.\ref{fig7}(a)). 
 We also note a sharp positive peak for the frequency domain of 2700-2800 cm$^{-1}$, 
 referring to free, dangling, or 
 weakly bound OD units (with D$_2$O not having four D-bonded neighbours),
  which are orientated away from the surface. For example, 
 O-D bonds pointing into vacuum are of this type. Note that
 positive intensities refer to dominantly ``upward'' and 
 negative intensities dominantly to ``downward'' orientations 
 of OD bonds.
 We also note that the intensity of this peak remains unchanged relative to the (unsupported) pure water/air interface computed in Ref.\cite{tdk2} (Fig.2 there),
 while the intensity of bonded, red-shifted 
  OD modes is relatively reduced as compared to this reference. This may imply an overall reduction of contributing bonded OD bonds, 
 relative to the free/non-bonded ones, in case of water on C/N surfaces compared to 
 the water/air interface. 
  This is in agreement with the frequency distribution of water molecules on g-C$_{3}$N$_{4}$ 
 as shown in Fig.\ref{fig1}(a), where the population of free/dangling OD modes is clearly 
 visible as a high-frequency shoulder.
%, 
% in contrast to bulk water,
% where it is absent\cite{tdk1,ojha}.
\\
 
Further, in Fig.\ref{fig7}(b) the second-order susceptibility 
 of  the water molecules on the C$_{2}$N surface is shown. The imaginary
 part of the second order susceptibility for this system is more 
 similar to the pure water-air interface\cite{tdk2}. Namely, 
 similar to the latter, 
 the  
 negative-intensity peak covering the spectral domain of 2200-2700 cm$^{-1}$ for the water/C$_{2}$N system is broader and of larger intensity 
 (comparable to the high-frequency peak) compared to the water/g-C$_{3}$N$_{4}$ surface. 
 (Note that in Refs.\cite{tdk1,tdk2}
 non-deuterated water was considered so absolute oxygen-hydrogen
 frequencies are different.)
% In simple words, the water/C$_2N$ system behaves more ``pure water-like'' 
% than the water/g-C$_3$N$_4$ system.
%We can thus infer that the orientation profile of water molecules has a more significant contribution of OD modes which are oriented away from the interface and 
% still shows strong hydrogen-bonding pattern. Further, the high-frequency 
% region corresponding to a frequency domain of 2700-2800 cm$^{-1}$ also has an overall strong intensity again referring to dangling/free OD modes of water molecules. \\
\\

To summarize, the vSFG spectra of OD modes on the g-C$_{3}$N$_{4}$ and C$_{2}$N surfaces  have similar features for the high frequency region around 2700-2800 cm$^{-1}$. But for the frequency region of 2200-2700 cm$^{-1}$, the spectral intensity is depleted in the case of g-C$_{3}$N$_{4}$, 
  while it is not for water/C$_{2}$N which behaves 
 similar to the water-air interface\cite{tdk1}. 
 We attribute this behaviour 
 to a propensity of the C$_2$N material to
 stabilize bonded OD bonds.
%This observation leads us to two possible
% scenarios for g-C$_3$N$_4$. Namely first,
% there could be a significant reduction of hydrogen bonds between the water molecules for g-C$_3$N$_4$ compared to C$_2$N, 
% due to which the spectral intensity shows a decrement. For the second scenario, the H- or surface-bonded OD modes are aligned more strongly 
% parallel to the surface 
% in case of the g-C$_3$N$_4$ surface and thus do not contribute to the vSFG spectrum. The observed spectrum 
% for g-C$_{3}$N$_{4}$ is most likely affected by both factors simultaneously being in action. 
%
\subsubsection{Time-resolved vSFG spectra}
Time-averaged vSFG spectra provide information about the orientation of molecular dipoles at the interface. Nevertheless, dynamics of these molecular dipoles  cannot be studied with the conventional vSFG technique.  Here, time-resolved vSFG spectra offer more insight.
 A time-dependent vibrational SFG spectrum can be obtained by 
 modifying Eq.(\ref{eq2}) as 
\begin{equation} \label{eq13}
{\chi^{2}_{abc}}\lvert_{\omega ({t}') = {\omega }'}  \hspace{0.4cm} =  \int e^{-i \omega t} \left \langle   {{  \alpha}_{ab}}({t}'+T_{w})   \cdot {{\mu}_{c}}({t}'+T_{w}+t)  \right \rangle
 dt \quad .
\end{equation}
Here, we compute the second-order susceptibility of a given vibrational oscillator provided that the oscillator was vibrating  at a frequency ${\omega}'$ at a  time instant ${t}'$. By systematically varying the value of the waiting time, 
 $T_{w}$ in Eq.(\ref{eq13}), the delay time between a pump pulse and the 
  overlapping IR/vis  
 pulses which generate an SFG signal, we can obtain frequency-resolved, time-dependent 
 susceptibilities and from there, TD-vSFG spectra\cite{tdk2}. 
 It is important to note that in the given computational framework, a priori information of the instantaneous frequency of the oscillators/OD modes is needed to obtain the TD-vSFG spectrum. This information is obtained 
 from the wavelet 
 transform as described above. Further, Eq.(\ref{eq13}) can easily be rewritten in the 
 ss-VVAF form (\ref{eq3}), which we use in the following to compute 
 Im $\chi^{(2)}(\omega; T_w)$ (actually, Im $\chi^{(2)}(\omega_{IR}; T_w, 
 \omega_{pump})$).
\\

\begin{figure}[!htpb]
\begin{center}

\includegraphics[width=1.2\textwidth]{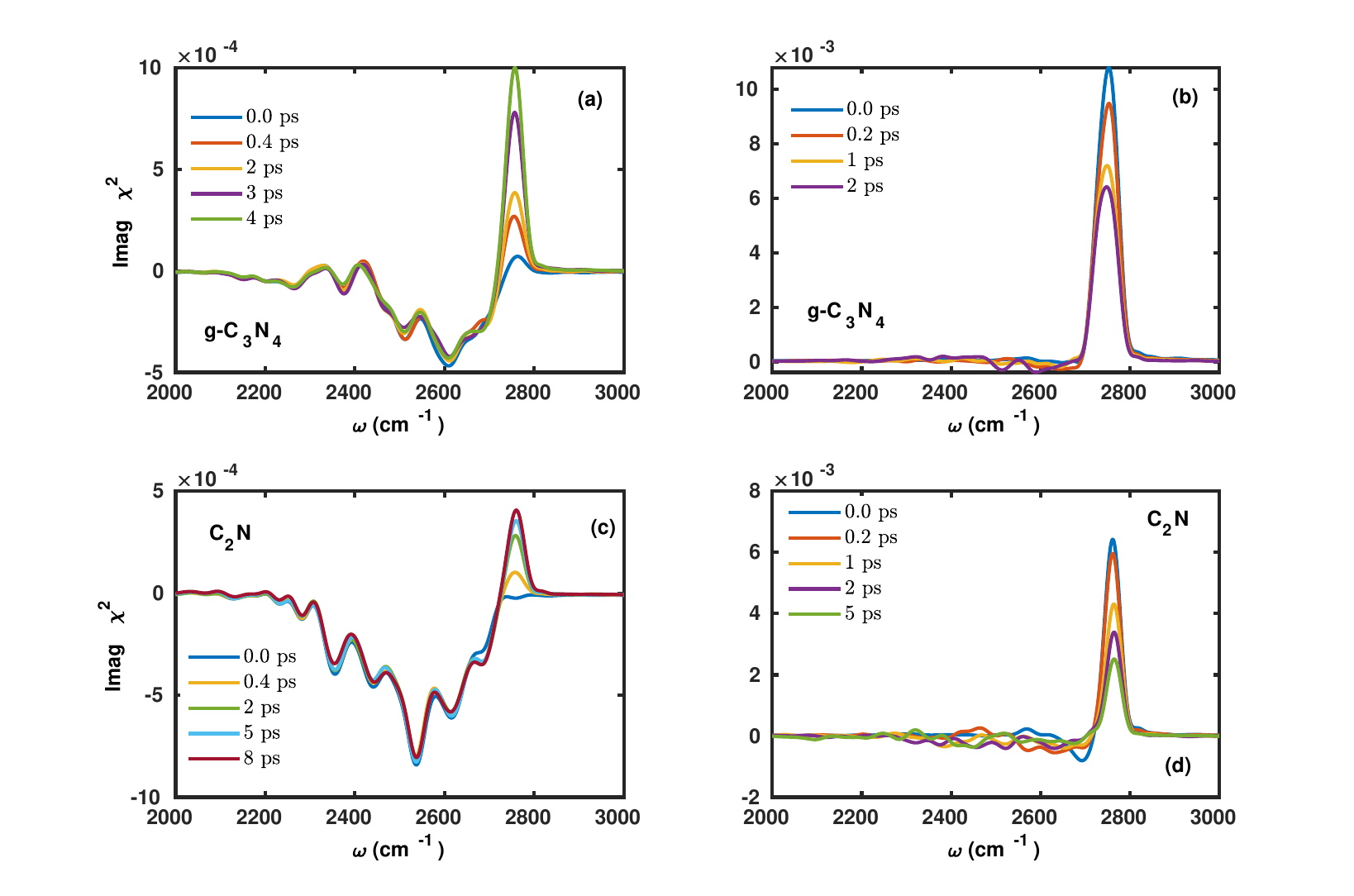}
\end{center}
\caption{\label{fig9} Imaginary part 
 of $\chi^{(2)}(\omega)$ at different delay times $T_w$, 
 computed for water on g-C$_{3}$N$_{4}$ for (a) the H-bonded OD modes and (b) the  free/dangling OD modes. For  C$_{2}$N, the same information 
 is given in panels (c) for bonded OD modes and (d) for free/dangling OD modes respectively. Now, for y-axes arbitrary but non-normalized 
 units are used to make the time-evolution of signal intensities visible. 
 } \end{figure}

 To calculate the TD-vSFG spectra of the hydrogen bonded OD modes, we have applied a broadband IR pump pulse corresponding to the frequency domain of 2200-2700 cm$^{-1}$. 
 Technically, this was done by computing the susceptibilities 
  of the water molecules based on their vibrational excitation frequency within this frequency range. 
The imaginary parts of the second-order susceptibility 
 of the hydrogen-bonded OD modes of the interfacial water molecules on 
 g-C$_{3}$N$_{4}$ corresponding to the waiting times $T_{w} = 0, 400, 2000, 3000 $ and $4000$ fs are shown in Fig.\ref{fig9}(a). The spectrum for the waiting time $T_{w} = 0$ fs, the ``impulsive limit'', 
  is dominated by a peak with negative intensity centered around 2600 cm$^{-1}$. With increasing waiting times, {\em e.g.} for $T_{w} =$ 400 fs,
  we see a formerly shallow shoulder in the high frequency region around 2750 cm$^{-1}$ 
 growing, corresponding to the free / dangling OD modes. Further, within 4.0 ps, the peak has attained the maximum intensity and becomes 
 saturated. The figure demonstrates the inter-conversion from hydrogen/surface-bonded 
 (``bound'') OD modes to free OD modes, taking place on this timescale.  \\

Similarly, we have calculated spectra of free / dangling OD modes 
 of g-C$_3$N$_4$ for waiting times $T_{w} = 0, 200, 1000 $  
 and 2000 fs in Fig.\ref{fig9}(b). Here we applied a narrowband IR pump pulse in the range of 2700-2900 cm$^{-1}$. 
 The spectrum for $T_{w} = 0$ fs is a single, positive
  peak centered around 2750 cm$^{-1}$. With increasing waiting times $T_{w} = 200, 1000, 2000$ fs, the frequency domain between 
 2200 and 2700 cm$^{-1}$ corresponding to the H-bonded region shows finite-valued, negative intensity, while at the same time the high-frequency, positive peak looses intensity. This indicates 
 interconversion of free OD modes to (H-)bound OD modes. 
 Closer inspection shows, as can also be seen from the 
 shorter timescale shown in Fig.\ref{fig9}(b), 
 that the free $\rightarrow$ bound interconversion 
 proceeds faster (within about 2 ps) in (b) 
 than the bound $\rightarrow$ free interconversion (about 4 ps) in (a).
 This is due to the free OD modes having a 
  strong propensity to form hydrogen bonds and 
 stabilize. 
%Therefore, the transition from free to bonded OD modes (b) is relatively faster compared to 
% the transition from H-bonded to free (a).
% In fact, for a waiting time beyond 1 ps, the intensity of bonded OD modes shows only marginal changes while the intensity of free OD modes keeps declining.
% It is important to note that the rate of inter-conversion is estimated based on the saturation of the peak intensity of the target frequency region which is the bonded OD modes for the TD-vSFG of the free OD modes and {\em vice versa}.  
 When compared to the pure water / air interface\cite{tdk2, tdk3}, 
 it is found that the hydrogen-bond inter-conversion dynamics for the bonded OD modes are overall slowed down whereas for the free OD, the timescale remains relatively unchanged in the presence of the g-C$_3$N$_4$ surface. 
\\

Along the same lines, we have also obtained the spectra of the bonded and dangling/free OD  bonds of water on the C$_{2}$N surface as shown in Fig.\ref{fig9}(c) and (d). The spectra of the bonded OD modes are shown for waiting times of $T_{w} =$0, 400, 2000, 5000 and 8000 fs, respectively. While the spectral features as seen in the TD-vSFG spectra are similar to previous calculations for the water / air interface\cite{tdk2},
  a key difference is a very slow temporal evolution of the peak for the free OD modes which extends up to nearly 8 ps. 
\\

The spectra of dangling/free OD modes are shown for the waiting times $T_{w} = $ 0, 200, 1000, 2000 and 5000 fs respectively. Again, the free OD modes have a strong propensity to form hydrogen bonds due to which the spectral domain corresponding to the bonded OD modes proceeds 
 mostly within 2 ps.  The 
 spectra of free OD modes of interfacial water molecules are interestingly independent of the surface, {\em i.e.}, g-C$_{3}$N$_{4}$ or C$_{2}$N. 
\\

To summarize, 
 we note that although the overall vibrational dynamics of OD modes on g-C$_{3}$N$_{4}$ and C$_{2}$N are similar 
 according to the frequency correlation function, Fig.\ref{fig5}(a) (with a tendency of faster H-bond dynamics for C$_2$N), the inter-conversion rates of free and bonded OD modes can be still significantly different (and of 
 other order). This refers in particular to 
 the bound $\rightarrow$ free interconversion of OD bonds, where D$_2$O/C$_2$N exhibits 
 a significantly longer ($\sim 8$ ps) timescale than 
 D$_2$O/g-C$_3$N$_4$ ($\sim 4$ ps). Here we want to  emphasize that the  time-resolved vSFG spectra of hydrogen-bonded OD modes on the g-C$_{3}$N$_{4}$/C$_{2}$N gives a comparative perspective of stabilization of the OD modes on  two surfaces.
The observed differences in the inter-conversion rates for the two systems imply that g-C$_{3}$N$_{4}$ could be more suitable for catalyzing interfacial 
  reactions involving water, {\em e.g.}, water 
 dissociation because the OD modes have a relatively  higher propensity to break the hydrogen bonds  ($\sim 4$ ps).   In contrast, C$_2$N may be more suited for processes like desalination or sieving which require stabilization of water 
 by the surface  with an average inter-conversion rate of nearly 8 ps .
 
  The key differences between both systems is that  
 the g-C$_{3}$N$_{4}$ has a buckled surface whereas C$_{2}$N is flat. Moreover, from the stoichiometric perspective g-C$_{3}$N$_{4}$ has a higher percentage of nitrogen as compared to C$_{2}$N. 
 These factors obviously affect the HB dynamics of adsorbed water. Finally  we note that the present inferences are drawn based on the TD-vSFG spectra of water molecules on a single monolayer of g-C$_{3}$N$_{4}$ or C$_{2}$N. The role of a multi-layered surface as well as density of  water molecules corresponding to liquid phase  in the simulation, may bring more insight into the catalytic properties of the two materials. Work along these lines is in progress in our laboratory.

\section{Summary} 
\label{sec4}
To summarize, we have studied the vibrational dynamics and (time-dependent) vSFG spectra of deuterated water molecules at g-C$_{3}$N$_{4}$ and C$_{2}$N surfaces respectively,  using DFT based AIMD 
 simulations. The most important results are as follows.
\begin{itemize}
\item
The time-averaged frequency distribution for the OD modes of water molecules on  both surfaces shows a high-frequency peak corresponding to free OD modes. 
% Since these peaks are not 
% seen at the water / air interface, 
%  this implies that these surfaces can facilitate the formation and stabilization  of free OD modes. It must also be said, however, that only thin water layers 
% were considered here ({\em cf.} Fig.\ref{fig1}), 
% so the expected ratio of ``free'' {\em vs.} ``bound''
% modes is comparatively large. 
 The ``bound'' OD bonds, either engaged in the 
 intermolecular H-bond network or interacting with the C/N surface, 
 are usually weakened and therefore red-shifted w.r.t. the high-frequency peak.
\item
The time-dependent decay of the OD frequency (measured by a frequency correlation function) is quite 
 similar for both surfaces with a timescale of about 4 ps which
  is considerably slower than that of the bulk liquid (deuterated) water.  
 Similar trends for the two studied C/N surfaces 
 arise when analyzing the H-bond network {\em via} a H-bond 
 measure $S_{HB}(t)$, which implies a slightly longer timescale for 
 g-C$_3$N$_4$ than for 
 C$_2$N.
\item 
 Time-independent vSFG spectra (measured by $\chi^2_{xxz}$), 
 indicate that the dangling OD bonds, pointing upward away from 
 the water / vacuum interface are dominant features in the spectra and are characterized by positive imaginary parts of $\chi^2_{xxz}$ at high frequencies.
 ``Bound'' OD modes are usually of lower intensity, have 
 negative imaginary parts of $\chi^2$ and are red-shifted.
 As a difference between the two surfaces, 
 the spectral domain around 2000-2600 cm$^{-1}$ of the vSFG spectrum for D$_2$O/C$_{2}$N is marked by a stronger negative intensity peak compared to g-C$_{3}$N$_{4}$. 
%Further, site-specific vSFG demonstrates that the OD modes make stronger bonds with N-sites as compared to C-sites.
\item
 Detailed insight into OD bond dynamics is provided by 
 TD-vSFG spectra, 
  with interesting trends for the two surfaces.
Although the vibrational frequency correlation functions are similar, the time-resolved vSFG of bonded OD modes saturates within about 4 ps for g-C$_{3}$N$_{4}$ whereas the same process takes nearly 8 ps for the C$_{2}$N surface. 
 The ``free'' $\rightarrow$ ``bound'' interconversion is 
 faster ($\sim 2$ ps) and more similar for both 
 surfaces. Thus  we can infer on the basis of slower interconversion rates for the bonded to free OD state that the C$_{2}$N surface has a propensity to stabilize  bonded OD moieties.
\end{itemize}
All of these observations imply different performance of porous C/N-\textcolor{red}{containing} materials in either catalysis or sieving. That key factors like 
 pore size, C/N ratio, or the three-dimensional structure of the 
 materials ({\em e.g.}, flat or buckled) influence the vibrational dynamics of 
 adsorbed water has been demonstrated, however, how these factors act 
 in detail is a matter of further research.
\section*{Acknowledgment}
P.S. acknowledges support by the Deutsche Forschungsgemeinschaft
 (DFG), through project Sa 547-18, and 
 Ch.P. 
 by Deutsche Forschungsgemeinschaft
 within Germany’s
Excellence Strategy -- 
EXC 2008/1-390540038, UniSysCat.
 The authors thank the Paderborn Center for Parallel Computing (PC$^2$) for computation time on supercomputer ``Noctua 2''.


\begin{thebibliography}{10}
\bibitem{michaelides}
%P. Ball, 
%\newblock {\em Life's Matrix: A Biography of Water}
%\newblock (Univ of California Press 2001).
{O. Björneholm, M.H. Hansen, A. Hodgson, L.-M. Liu, D.T. Limmer, A. Michaelides, P. Pedevilla, J. Rossmeisl, H. Shen, G. Tocci, E. Tyrode, M.-M. Walz, J. Werner, and H. Bluhm},
\newblock {\em  Chem. Rev.} 2016, {\bf 116}, 7698-7726.

\bibitem{marcus}
Y. Jung and R. A. Marcus
\newblock {\em J. Amer. Chem. Soc.} 2007, { \bf 129}, 5492–5502.
\bibitem{skinner}
J. L. Skinner, P. A. Pieniazek, and S. M. Gruenbaum,
\newblock {\em Acc. Chem. Res.} 2012, { \bf 45}, 93-100.

\bibitem{pavel}
P. Jungwirth,
\newblock {\em J. Chem. Phys. Lett.} 2015, { \bf 6}, 2449–2451.

\bibitem{gonella}
G. Gonella et al.,
\newblock {\em Nat. Rev. Chem.} 2021, { \bf 5}, 466–485.

\bibitem{zhu}
J. Zhu et al.
\newblock {\em Angew. Chem., Int. Ed.}  2015,
{\bf 54}, 9111-9114.

\bibitem{nicolas}
N. G. Hörrmann, Z. Guo, F. Ambrosio, O. Andreussi, A. Pasquarello, and 
 N. Marzari, 
\newblock {\em npj Comput Mater} 2019,
{\bf 5}, 100-105.

\bibitem{anton1} 
X. Wang, K. Maeda, A. Thomas, K. Takanabe, G. Xin, J. M. Carlsson, K. Domen, and M. Antonietti , 
\newblock {\em Nat. Mater.} 2009,  {\bf 8} , 76-80.  

\bibitem{anton2} 
Y. Wang, X. Wang, and M. Antonietti,
\newblock {\em Angew. Chem. Int. Ed} 2009 , {\bf 51}, 68-89.


\bibitem{anton3} 
X. Wang , S. Blechert and M. Antonietti,
\newblock {\em ACS Catal.} 2012,  {\bf 2}, 1596-1606, .

\bibitem{walczak2018}
R. Walczak, B. Kurpil, A. Savateev, T. Heil, J. Schmidt, Q. Qin, M. Antonietti, and M. Oschatz, 
\newblock {\em Angew. Chem. Int. Ed.} 2018, {\bf 57}, 10765-10770.


\bibitem{oschatz}
J. Heske, R. Walczak, J.D. Epping, S. Youk, S.K. Sahoo, M. Antonietti, T.D. K\"uhne, M. Oschatz, 
\newblock {\em J. Mat. Chem. A} 2021, {\bf 9}, 22563-22572.

\bibitem{penschke}
Ch. Penschke, A. Zehle, N. Jahn, R.E. von Zander, R. Neumann, A. Beqiraj, and P. Saalfrank,
%          {\em Water on Porous, Nitrogen-containing Two-dimensional Carbon Materials: The Performance of Computational Model Chemistries.}
\newblock {\em Phys. Chem. Chem. Phys.} 2022, {\bf 24}, 14709-14726.

\bibitem{grueneis}
T. Sch{\"a}fer, A. Gallo, A. Irmler, F. Hummel, and A. Gr{\"u}neis,
\newblock {\em J. Chem. Phys.} 2021, 155, 244103.

\bibitem{wu} 
 H.-Z. Wu,   L.-M. Liu, S.-J. Zhao
\newblock { Phys. Chem. Chem. Phys.}, 2014,  { \bf 16}, 3299-3304.

\bibitem{kasai} 
 S. M. Aspera1, M. David, and H. Kasai 
 \newblock { Jpn. J. Appl. Phys}  2010,  { \bf 49}, 115703.
 
 \bibitem{zou}
 X. Zou, M. Li, S. Zhou, Ch. Chen, J. Zhong, A. Xue, Y. Zhang, and Y. Zhao,
  \newblock {J. Membr. Sci.}  2019,  { \bf 585}, 81-89.
  
  \bibitem{peter}
J. Wirth, R. Neumann, M. Antonietti and P. Saalfrank, 
\newblock {\em Phys. Chem. Chem. Phys} 2014, {\bf 16}, 15917-15926.

\bibitem{hummer}
 I.-C. Yeh and G. Hummer, 
\newblock {\em J. Phys. Chem. B} 2004, {\bf 108}, 15973-15879.

%\bibitem{fardis}
%M. Fardis, M. Karagianni, L. Gkoura, and G. Papavassiliou,
% G. Self-Diffusion in Confined Water: A Comparison between the Dynamics of Supercooled Water in Hydrophobic Carbon Nanotubes and Hydrophilic Porous Silica.
%Int. J.  Mol.  Sci. 2022, {\bf 23}, 14432.

\bibitem{netz}
 N. Kavokine, R. R. Netz, and L. Bocquet, 
%Fluids at the Nanoscale: From Continuum to Subcontinuum
%Transport, 
 {\em Annu. Rev. Fluid Mech.} 2021, {\bf 53}, 377–410.

% \bibitem{tdk6}
% J. Heske et al.,
% \newblock {\em J. Mater. Chem. A} 2021, {\bf 9}, 22563-22572.
 
%  \bibitem{kishore}
% M.R.A. Kishore and P. Ravindran,
%  \newblock {\em Chem. Phys. Chem} 2017, {\bf 18}, 1526 –1532.

% \bibitem{tianb}
% B. L. He,   J. S. Shena  and  Z. X. Tianb,
%   \newblock {\em  Phys. Chem. Phys. Chem} 2016, {\bf 18}, 24261-24269.
   
% \bibitem{yang} 
%Y. Yang, W. Li, H. Zhou, X. Zhang and  M. Zhao ,
%\newblock {\em Sci. Rep. } 2016, {\bf 6}, 29218.
    

\bibitem{shen1}
 Q. Du, R. Superfine, E. Freysz, and   Y. R. Shen, 
%Vibrational spectroscopy of water at the vapor/water interface.
\newblock {\em  Phys. Rev. Lett.}, 1993, {\bf 70}, 2313--2316.

\bibitem{shen2}
 Y. R. Shen,   and V. Ostroverkhov,
%Sum-Frequency Vibrational apectroscopy on Water Interfaces: Polar Orientation of Water Molecules at Interfaces.
\newblock {\em  Chem. Rev.}, 2006, {\bf 106}, 1140--1154.

\bibitem{shen3}
V. Ostroverkhov, G. A. Waychunas, Y. R.  Shen, 
%New information on water interfacial structure revealed by phase-sensitive surface spectroscopy.
\newblock {\em  Phys. Rev. Lett.}, 2005, {\bf 94}, 046102.

\bibitem{morita1}
 A. Morita and J. T.  Hynes, 
%A theoretical analysis of the sum frequency generation spectrum of the water surface.
\newblock {\em Chem. Phys.} 2000 {\bf 258}, 371--390.

\bibitem{morita2}
A. Morita and J. T.  Hynes,
%A Theoretical Analysis of the Sum Frequency Generation Spectrum of the Water Surface. II. Time-Dependent Approach.
\newblock {\em J. Phys. Chem. B}, 2002, {\bf 106}, 673--685.

\bibitem{morita3}
T. Ishiyama, T.  Imamura and  A. Morita, 
%Theoretical Studies of Structures and Vibrational Sum Frequency Generation Spectra at Aqueous Interfaces.
\newblock {\em  Chem. Rev.},  2014, {\bf 114}, 8447--8470.

\bibitem{morita-book}
 A. Morita,
Theory of Sum Frequency Generation Spectroscopy
(Springer, Singapore 2018)

\bibitem{tdk1}
D. Ojha and T. D.  K\"uhne,
\newblock {\em Molecules}, 2020, {\bf 25}, 3939.

 
\bibitem{tr-sfg}
 E. H. G. Backus, J. D. Cyran, M. Grechko, Y. Nagata, and M. Bonn,
 \newblock {\em J. Phys. Chem. A}  2018, {\bf 122}, 2401–2410.
 
    

\bibitem{tdk2}
D. Ojha, N. K. Kaliannan,  T. D. K\"uhne
%Time-dependent vibrational sum-frequency generation spectroscopy of the air-water interface
\newblock {\em Commun. Chem.}, 2019, { \bf 2},  116.

\bibitem{tdk3}
D. Ojha and T. D.  K\"uhne,
\newblock {\em Sci. Rep.}, 2021, {\bf 11}, 2456.


 \bibitem{yuki}
T. Ohto, K. Usui, T.  Hasegawa,  M. Bonn, and Y. Nagata,
%Toward ab initio molecular dynamics modeling for sum-frequency generation spectra; an efficient algorithm based on surface-specific velocity-velocity correlation function.
\newblock {\em J. Chem. Phys.}, 2015, {\bf 143}, 124702.


\bibitem{vasp1}
G. Kresse and  J. Furthmüller,
\newblock {\em Comput. Mater. Sci.} 1996, { \bf 6}, 15–50.

\bibitem{vasp2}
G. Kresse and J. Furthmüller,
\newblock {\em Phys. Rev. B} 1996, { \bf 54}, 11169–11186.

\bibitem{blochl}
P. E. Blöchl,
\newblock {\em Phys. Rev. B} 1994, { \bf 50}, 17953–17979.

\bibitem{psp}
G. Kresse and D. Joubert,
\newblock {\em Phys. Rev. B}   1999, { \bf 59}, 1758–1775.

\bibitem{pbe}
J. P. Perdew, K. Burke and  M. Ernzerhof,
\newblock {\em Phys. Rev. Lett.}  1996, { \bf 77}, 3865–3868.


\bibitem{dftd}
S. Grimme, J. Antony, S. Ehrlich and  H. Krieg,
\newblock {\em J. Chem. Phys.} 2010, { \bf 132}, 154104.

\bibitem{dftd2}
S. Grimme, S. Ehrlich and  L. Goerigk,
\newblock {\em J. Comput. Chem.}  2011,  { \bf 32}, 1456–1465.

\bibitem{michal2}
M.J. Gillan, D. Alf{\'e}, and A. Michaelides, 
\newblock {\em J. Chem. Phys.}  2016,  { \bf 144}, 13091.

\bibitem{nose}
S. Nos{\'e},
\newblock {\em J. Chem. Phys.} 1984, { \bf 81}, 511–519.

\bibitem{melani2}
G. Melani, Y. Nagata, R.K. Campen, and P. Saalfrank,
%\newblock Vibrational spectra of dissociatively
%adsorbed D2O on Al-terminated -
%Al2O3(0001) surfaces from ab initio
%molecular dynamics
\newblock {\em J. Chem. Phys.}, 2019, {\bf 150}, 244701.


\bibitem{wavelet}
A. Semparithi  and  S. Keshavamurthy,
\newblock { Phys. Chem. Chem. Phys.}, 2003,  { \bf 5}, 5051--5062.

\bibitem{tdk4}
D. Ojha, A. Henao, and T. D.  K\"uhne, 
%Nuclear quantum effects on the vibrational dynamics of liquid water.
\newblock {\em J. Chem. Phys.}, 2018, {\bf 148}, 102328.

\bibitem{tdk5}
D. Ojha, K. Karhan, and T. D.  K\"uhne, 
%\newblock On the hydrogen bond strength and vibrational spectroscopy of liquid water.
\newblock {\em Sci. Rep.}, 2018, {\bf 8}, 16888.

\bibitem{ojha}
D. Ojha and T.D. K\"uhne,
\newblock {\em Phys. Chem. Chem. Phys.}, 2023, {\bf 25}, 13442.


\bibitem{melani}
G. Melani, Y. Nagata, J. Wirth, and P. Saalfrank, 
%\newblock Vibrational spectroscopy of hydroxylated -Al2O3(0001) surfaces with and without
%water: An ab initio molecular dynamics study
\newblock {\em J. Chem. Phys.}, 2018, {\bf 149}, 014707.

\bibitem{marx}
M. Heyden, J. Sun, S. Funkner, G. Mathias, H. Forbert, M. Havenith, and D. Marx
Proc. Nat. Acad. Sci., 2010,  { \bf 107}, 12068--12073.


\bibitem{chandra} 
D. Ojha and A. Chandra,
Chem. Phys. Lett. 2020, { \bf 751}, 137493.


\bibitem{chandra2}
J.R. Choudhuri and A. Chandra,
J. Chem. Phys. 2014, { \bf 141}, 194705.




















\end{thebibliography}
\end{document}